
\input harvmac
\input epsf             

\catcode`\@=11      
\def\lsim{\mathrel{\mathpalette\@versim<}}
\def\gsim{\mathrel{\mathpalette\@versim>}}
\def\@versim#1#2{\vcenter{\offinterlineskip\ialign
    {$\m@th#1\hfil##\hfil$\crcr#2\crcr\sim\crcr}}}

\def\theorem#1.#2\par{\medbreak\noindent{\bf#1.\enspace}
						{\sl#2}\par\medbreak}
\def\vacv #1{\ifmmode \langle 0 \mid #1 \mid 0 \rangle
		\else $ \langle 0 \mid #1 \mid 0 \rangle $ \fi}

\def\sec{\ifmmode \,\, {\rm sec} \else sec \fi}
\def\eV {\ifmmode \,\, {\rm eV} \else eV \fi}
\def\keV{\ifmmode \,\, {\rm keV} \else keV \fi}
\def\MeV{\ifmmode \,\, {\rm MeV} \else MeV \fi}
\def\GeV{\ifmmode \,\, {\rm GeV} \else GeV \fi}
\def\TeV{\ifmmode \,\, {\rm TeV} \else TeV \fi}
\def\pbarn{\ifmmode \,\, {\rm pb} \else pb \fi}

\def\QEDBox{$\,$\raise .5ex\hbox{\vrule height3pt width5pt depth2pt}}

\def\bigabs#1{\ifmmode \,\biggm|\! #1 \!\biggm|\,
		\else $ \,\biggm|\! #1 \!\biggm|\, $ \fi}
\def\abs#1{\ifmmode \,\mid\! #1 \!\mid\,
		\else $ \,\mid\! #1 \!\mid\, $ \fi}
\def\norm#1{\ifmmode \,\mid\!\mid\! #1 \!\mid\!\mid\,
		\else $ \,\mid\!\mid\! #1 \!\mid\!\mid\,$ \fi}

\def\comment#1{ }
\def\myinstitution{\vskip 20pt
   \centerline{\it Enrico Fermi Institute and Department of Physics }
   \centerline{\it University of Chicago, Chicago, IL 60637 }
   \vskip .3in
}
\def\myemail{\footnote{$^\ast$}
	{e-mail : tonyg@yukawa.uchicago.edu}}

\Title{\vbox{\hbox{EFI-93-37} \hbox{cond-mat/9307010}}}
      {\vbox{\centerline{Feedback Effects in Superconductors}}}

\centerline{Tony Gherghetta\myemail {\hskip 3pt} and
             {\hskip 3pt} Yoichiro Nambu}
\myinstitution


\vskip .3in

\noindent
We calculate corrections to the BCS gap equation caused by the
interaction of electrons with the collective phase and amplitude modes
in the superconducting state. This feedback reduces the BCS gap
parameter, $\Delta$, and leaves the critical temperature, $T_c$,
unchanged. The feedback effect is proportional to $(\Delta /
{\scriptstyle{\cal E}_F})^2$, where ${\scriptstyle{\cal E}_F}$ is the
Fermi energy. This is a negligible correction for type-I
superconductors. However in type-II superconductors the feedback
effect is greatly enhanced due to smaller Fermi velocities, $v_F$, and
may be responsible for effects seen in recent experimental data on
organic superconductors.

\Date{June 93}

\lref\ampref{R.~Sooryakumar and M.~V.~Klein, Phys.~Rev.~Lett. {\bf 45},
         660 (1980)
      \semi Phys.~Rev. {\bf B23}, 3213 (1981);{\bf B23}, 3223 (1981).}
\lref\andbog{P.~W.~Anderson, Phys.~Rev. {\bf 110}, 827 (1958);
            {\bf 112}, 1900 (1958)
      \semi N.~N.~Bogoliubov, Nuovo Cimento, {\bf 7}, 794 (1958).}
\lref\organicref{K.~I.~Pokhodnia, A.~Graja, M.~Weger and D.~Schweitzer,
            Z.~Phys.~B, {\bf 90}, 127 (1993).}
\lref\frohlich{H.~Fr\"ohlich, Phys.~Rev. {\bf 79}, 845 (1950).}
\lref\schrieffer{J.~R.~Schrieffer, {\it Theory of Superconductivity},
      W.~A.~Benjamin Inc, New York, 1964.}
\lref\nambu{Y.~Nambu, Phys.~Rev. {\bf 117}, 648 (1960).}
\lref\njl{Y.~Nambu and G.~Jona-Lasinio Phys.~Rev. {\bf 122}, 345 (1961).}
\lref\bcs{J.~Bardeen, L.~N.~Cooper and J.~R.~Schrieffer, Phys.~Rev.
       {\bf 108}, 1175 (1957).}
\lref\lurie{S.~Cremer, M.~Sapir and D.~Lurie, Nuovo Cimento, {\bf B6},
        179 (1971).}
\lref\goldstone{J.~Goldstone, Nuovo Cimento, {\bf 19}, 154 (1961).}
\lref\lv{P.~B.~Littlewood and C.~M.~Varma, Phys.~Rev. {\bf B26}, 4883 (1982)
      \semi C.~A.~Balseiro and L.~M.~Falicov, Phys.~Rev.~Lett. {\bf 45}, 662
      (1980).}
\lref\kresin{V.~Z.~Kresin and S.~A.~Wolf, {\it Fundamentals of
        Superconductivity}, Plenum Press, New York, 1990}

\nfig\bubblefig{The infinite sum of bubble diagrams for the quasiparticle-
       quasiparticle scattering amplitude. The $\tau_i$ represents either
       $\tau_1$ ($\sigma$ mode) or $\tau_2$ ($\pi$ mode).}
\nfig\sigmafig{The quasiparticle self-energy diagrams arising from the
      $\sigma$ mode coupling, where (a) depicts the tadpole diagram
      and (b) shows the Weisskopf term.}
\nfig\solnfig{Comparison of the solution of the gap equation with feedback
      effect (dashed line) with the normal BCS result (solid line) for
      typical type-II superconductor parameter values.}

In the BCS theory of superconductivity \bcs, there exist two distinct
collective modes corresponding to the fluctuations of the phase and
amplitude of the superconducting gap. The phase or Anderson-Bogoliubov
mode \andbog\ has been known for a very long time to be important in
maintaining gauge invariance in the BCS theory \nambu. In the presence
of a Coulomb field, the phase mode, $(\pi)$ interacts strongly with
the Coulomb field to become the plasmon mode. On the other hand the
amplitude mode, $(\sigma)$ is unaffected by Coulomb interactions, so
that this mode remains intact. This decoupling feature of the
amplitude mode means that it is not easily observable and it was
only recently that such a mode was discovered in the charge density
wave compound ${\rm NbSe}_2$ through the coupling to long-wavelength
phonons \refs{\ampref{--}\lv}.

In this Letter we wish to consider the effects of these collective
modes back on the superconducting state. In the effective four-Fermi
interaction BCS theory, an effective coupling between the collective
modes and the quasiparticles induces self-energy corrections to the
quasiparticle propagator. These corrections can either enhance the
attraction between Cooper pairs and so contribute positively to the
superconducting state, or they can act negatively on the
superconducting state and reduce the gap parameter $\Delta$. The
magnitude of these corrections is proportional to $(\Delta/
{\scriptstyle{\cal E}_F})^2$. These are negligible corrections for
type-I superconductors where typically $\Delta/{\scriptstyle{\cal
E}_F}\sim 10^{-3}$. However the feedback effects may become important
if the typical Fermi energies are much smaller. This is the case in
type-II superconductors where $v_F\simeq 10^{6} {\rm cm\,s^{-1}}$.
Recent experiments in organic superconductors \organicref, where
typical Fermi energies are small, hint at the possibility that such a
scenario may be at work. We will now present a calculation of these
corrections and show how the superconducting state is affected.

Let us first recall some basic features of the field theoretic
formulation of BCS superconductivity \refs{\schrieffer{--}\lurie}.
In the BCS ansatz \bcs\ the Fr\"ohlich effective electron-electron
interaction \frohlich\ is replaced by a contact potential
\eqn\potential{
	V({\vec x}-{\vec x}^\prime)=-V\delta^3({\vec x}-{\vec x}^\prime)
}
where $V>0$. The effective Lagrangian is given by
\eqn\namlag{
            {\cal L}= i\Psi^\dagger {\dot \Psi} - \Psi^\dagger
            {\scriptstyle{\cal E}} \tau_3\Psi + \half V
            \Psi^\dagger\tau_3\Psi\Psi^\dagger\tau_3\Psi
}
where ${\scriptstyle{\cal E}}$ is the electron kinetic energy measured
from the Fermi energy and we have used the two-component notation \nambu
\eqn\twocomp{
	\Psi=\left(\matrix{\psi_\uparrow  \cr
                           \psi_\downarrow^\dagger }
             \right)
}
to represent the Bogoliubov-Valatin fermionic quasiparticle modes.
In the superconducting state the Lagrangian \namlag\ is written as a sum
of a free term ${\cal L}_0$ plus an interaction piece
${\cal L}_I$
\eqn\sumlag{\eqalign
	  {{\cal L}_0 &= i\Psi^\dagger {\dot \Psi} - \Psi^\dagger
          {\scriptstyle{\cal E}}\tau_3\Psi - \Delta
           \Psi^\dagger\tau_1\Psi \cr
	   {\cal L}_I &=\half V \Psi^\dagger \tau_3 \Psi \Psi^\dagger
	   \tau_3 \Psi +\Delta \Psi^\dagger \tau_1 \Psi}
}
where we have introduced the mass gap $\Delta$. The bare quasiparticle Green's
function corresponding to ${\cal L}_0$ is
\eqn\greenfn{
        G(k)=i{k^0 1+{\scriptstyle{\cal E}} \tau_3 +\Delta \tau_1 \over
            (k^0)^2-E^2+i\epsilon}
}
where $E^2={\scriptstyle{\cal E}}^2+\Delta^2$ is the quasiparticle
excitation energy. In ${\cal L}_I$ we have to ensure that there are no
self energy corrections proportional to $\tau_1$ in order to maintain
consistency with the ansatz that ${\cal L}_0$ describes the superconducting
ground state with mass gap $\Delta$. Using a Fierz identity for the Pauli
matrices, \lurie\ this leads directly to the BCS gap equation \bcs
\eqn\gapeqn{
           1=\half V \int {d^3p\over (2\pi)^3}{1\over
           \sqrt{{\scriptstyle{\cal E}}^2+\Delta^2}} \equiv J(\Delta),
}
where the integral is cutoff at the Debye energy, $\omega_D$.

To exhibit the collective modes of the superconducting state, let us
examine the quasiparticle-quasiparticle scattering amplitude generated
by the infinite sum of bubble diagrams, as shown in \bubblefig. The
scattering amplitude is a simple geometric series and is easily summed
to give
\eqn\matrixel{
	   {\cal A}_{\pi,\sigma}(k)={-{\textstyle{i\over 2}} V
           \over 1- I_{\pi,\sigma}(k)}
}
where
\eqn\pimode{
	    I_\pi(k)=-i\,\half V\int{d^4 p\over (2\pi)^4}\,{\rm Tr}\left[
            \tau_2 G(p+{k\over 2}) \tau_2 G(p-{k\over 2})\right]
}
and
\eqn\sigmamode{
	   I_\sigma(k)=-i\,\half V\int{d^4 p\over (2\pi)^4}\,{\rm Tr}\left[
            \tau_1 G(p+{k\over 2}) \tau_1 G(p-{k\over 2})\right]
}
are the integrals for the two types of single bubble diagrams. The poles
of the scattering amplitude \matrixel\ occur when $I_{\pi,\sigma}(k)=1$.
At zero momentum transfer (${\vec k}=0$), the integrals \pimode\ and
\sigmamode\ can be written in the form
\eqn\newpi{
          I_\pi(k_0^2=\omega^2)=1-\half V \int{d^3 p\over (2\pi)^3}
          \,{1\over E_p}\,{\omega^2\over \omega^2-4 E_p^2}
}
\eqn\newsigma{
	I_\sigma(k_0^2=\omega^2)=1-\half V \int{d^3 p\over (2\pi)^3}
        \,{1\over E_p}\,{\omega^2-4\Delta^2\over \omega^2-4 E_p^2}
}
where we have used the BCS gap equation \gapeqn. It is then obvious that
$I_\pi(\omega^2=0)=I_\sigma(\omega^2=4\Delta^2)=1$ and the
quasiparticle-quasiparticle scattering amplitude has poles at $\omega^2=0$
and $\omega^2=4\Delta^2$ which represent the phase ($\pi$) and amplitude
($\sigma$) modes respectively \njl. For nonzero momentum transfers $\vec k$,
we can Taylor expand the integrands in \pimode\ and \sigmamode\ to obtain
the dispersion relations for the collective modes \lv\
\eqn\pidisprel{
       E_\pi^2({\vec k})={\textstyle{1\over 3}}v_F^2 {\vec k}^2
}
\eqn\sigdisprel{
       E_\sigma^2({\vec k})=4\Delta^2+{\textstyle{1\over 3}}v_F^2 {\vec k}^2
}
where $v_F$ is the Fermi velocity.

The effective quasiparticle-collective mode coupling is obtained from the
residue at the pole of the scattering amplitude \matrixel. Using \newpi, the
quasiparticle-$\pi$ mode coupling is
\eqn\picpdefn{
          f_\pi^2=-\half V \left( {d I_\pi\over d \omega^2}
          \biggl.\biggr|_{\omega^2=0} \right)^{-1}
	  {\textstyle{=\atop {\scriptstyle{\omega_D \gg \Delta}}}}
          {4 \Delta^2 \over N({\scriptstyle{\cal E}_F})}
}
where $N({\scriptstyle{\cal E}_F})= {m k_F\over \pi^2}$ is the density
of states at the Fermi surface. If we attempt a similar procedure for
the $\sigma$ mode then it turns out that the corresponding integral in
\picpdefn\ is divergent, because the pole coincides with the two-particle
threshold. This is the inadequacy of modelling the BCS theory by the
$\sigma$ model. We will simply circumvent this problem by assuming
$f_\sigma=f_\pi$ as in the Ginzburg-Landau theory. This is a good
approximation in the weak coupling limit.

What are the effects of the collective modes on the quasiparticle
self-energy?  First, we assume that a Coulomb field is present, so
that the Goldstone $\pi$ mode turns into the massive plasmon mode
\schrieffer. In order to correctly take into account the plasmon mode,
we need to start with the original Coulomb and phonon interactions
instead of the effective four-Fermi interaction, $V$. This is beyond the
scope of this Letter, so we will ignore its effects on the quasiparticle
self-energy. However we expect this contribution to be small because
the plasmon mass is large compared to $\Delta$ and the Debye energy,
$\omega_D$.

For the massive $\sigma$ mode there will be two contributions
to the quasiparticle self-energy. The first contribution comes from
the tadpole term shown in \sigmafig. It is given by
\eqn\tadpole{
	\Sigma_t=-i {\Delta J(\Delta) \over 1- I_\sigma(0)}
}
However this term is already implicitly included in the BCS gap equation
and its inclusion would amount to a double counting of diagrams. To see
this more clearly, consider the effect of adding a small bare term
$\Delta_0$ to the gap equation \gapeqn
\eqn\bareterm{
	\Delta=\Delta_0+\Delta J(\Delta)
}
If we now seek a perturbative solution of \bareterm\ of the form
$\Delta+\delta\Delta$ then we obtain
\eqn\deltadelta{\eqalign{
	\delta\Delta&=\Delta_0+{\partial (\Delta J(\Delta))\over
	\partial\Delta} \delta\Delta \cr
	&={\Delta_0 \over 1- I_\sigma(0)}}
}
Thus comparing \tadpole\ and \deltadelta\ we see that the tadpole term
appears only as the response of $\Delta$ to a nonzero $\Delta_0$.

The second contribution results from contracting the crossed tree
diagram, leading to the ``Weisskopf term'' shown in \sigmafig. This
contribution will act negatively on the superconducting state because
contracting the crossed diagram involves a sign change. Hence the Weisskopf
term will act to reduce the gap $\Delta$. In order to calculate the
Weisskopf term we reinterpret the quasiparticle-quasiparticle scattering
amplitude, ${\cal A}_\sigma$ as arising from the exchange of the
$\sigma$ mode with propagator
\eqn\sigmaprop{
       G_\sigma(k)=i{1\over (k^0)^2-E_\sigma^2({\vec k})}
}
where we have approximated the continuum cut solely by the $\sigma$ mode pole.
This considerably simplifies the equations and the corrections arising from
the continuum contribution in \matrixel\ do not affect the qualitative
behaviour. The Weisskopf self-energy term may now be written as
\eqn\weissint{
        \Sigma_W(k)=i\,f_\pi^2 \int {d^4p\over (2\pi)^4}
        \, \tau_1 G(p) G_\sigma(p-k)\tau_1
}
and we will evaluate it at the Fermi surface : $k_0=\Delta$,
$\abs{\vec k}=k_F$. The term proportional to $\tau_1$ which gives a
contribution to the gap in the limit ${\scriptstyle{\cal E}_F}\gg\omega_D
\gg\Delta$ is
\eqn\weisstau{\eqalign{
	\delta Z_\Delta = {\textstyle{3\over 8}} \left({\Delta\over
	{\scriptstyle{\cal E}_F}}\right)^2 &\int_0^{\omega_D}
	d{\scriptstyle{\cal E}} {1\over\sqrt{{\scriptstyle{\cal E}}^2
	+\Delta^2}} \cr &\times \ln \left[{\textstyle{1\over 4}}\left(
	{{\scriptstyle{\cal E}}\over{\scriptstyle{\cal E}_F}}\right)^2
	+{3\over 8{\scriptstyle{\cal E}_F^2}} \sqrt{\left({\scriptstyle{\cal
	E}}^2+\Delta^2 \right)\left( {\textstyle{1\over 3}} {\scriptstyle{\cal
	E}}^2 +4\Delta^2\right)} +{\textstyle{3\over
	4}}\left({\Delta\over{\scriptstyle{\cal E}_F}} \right)^2\right]}
}
where $\omega_D$ is the Debye frequency cutoff.

The Weisskopf term will also give corrections to the $\tau_3$ and $\bf 1$
terms in ${\cal L}_0$. The corrections to $\tau_3$ will renormalize the
chemical potential and the electron mass and give rise to an effective
electron mass, $m^\ast$. The term proportional to the identity matrix,
$\bf 1$, adds a contribution $k_0(\delta Z_\Psi)$ to the energy $k_0$.
Defining $Z_\Psi=1-\delta Z_\Psi$, this corresponds to a wavefunction
renormalization $\Psi\rightarrow Z_\Psi^{-1/2} \Psi$ and modifies the
mass gap term by $Z_\Psi^{-1}$. Evaluating the wave function
renormalization constant at the Fermi surface in the limit
${\scriptstyle{\cal E}_F}\gg\omega_D\gg\Delta$ gives
\eqn\waveren{
	\delta Z_\Psi = -{3\over 8} \left({\Delta\over
        {\scriptstyle{\cal E}_F}}\right)^2 \left[\left(3-\sqrt{3}\right)
        \ln {2\omega_D \over \Delta} - 0.844
        +{\cal O}({\omega_D\over{\scriptstyle{\cal E}_F}})\right].
}
Thus the total self-energy contributions to the gap arising from the
Weisskopf term will be $\Delta_W=\Delta(\delta Z_\Delta + \delta Z_\Psi)$
where we have kept terms to lowest order in the correction parameter
$(\Delta /{\scriptstyle{\cal E}_F} )^2$. Thus the BCS gap equation with
the Weisskopf corrections in the limit ${\scriptstyle{\cal E}_F}\gg\omega_D
\gg\Delta$ is
\eqn\newgap{\eqalign{
        1=\half V N({\scriptstyle{\cal E}_F})\ln{2\omega_D\over\Delta}
	&-{3\over 8}\left({\Delta\over{\scriptstyle{\cal E}_F}}
        \right)^2 \left[\left(\ln{2\omega_D\over\Delta}\right)^2 +
        \ln{2\omega_D\over\Delta}\left(2\ln{{{\scriptstyle{\cal E}_F}}
	\over\omega_D}+0.762\right) -2.389\right] \cr
	&-{3\over 8} \left({\Delta\over{\scriptstyle{\cal E}_F}}\right)^2
        \left[\left(3-\sqrt{3}\right)\ln {2\omega_D\over\Delta}
	-0.844\right]}
}
where we have evaluated the integral in \weisstau\ to ${\cal O}(\ln x/x^2)$
where $x=\omega_D/\Delta$. In normal type-I superconductors $(\Delta
/{\scriptstyle{\cal E}_F})^2 \sim 10^{-6}$ because $v_F \simeq 10^8
{\rm cm\,s}^{-1}$. This is quite a small correction compared to
$V N({\scriptstyle{\cal E}_F})\sim 0.25$. However in type-II
superconductors the Fermi velocity is smaller: $v_F \simeq
10^6 {\rm cm\,s}^{-1}$, and the gap parameter is larger, so that the
correction, $(\Delta /{\scriptstyle{\cal E}_F} )^2 \sim 10^{-2}$.
Note that this does not contradict the fact that we assumed ${\scriptstyle
{\cal E}_F}\gg\omega_D\gg\Delta$, because the corrections are always
proportional to $(\Delta/{\scriptstyle{\cal E}_F})^2$. However we
need to obtain the coefficients of $(\Delta /{\scriptstyle{\cal E}_F} )^2$
in the limit ${\scriptstyle{\cal E}_F}\gsim\omega_D$. We present the full
exact expressions for $\delta Z_\Delta$ and $\delta Z_\Psi$ below.

All the above results are for $T=0$. The results at finite temperature
are obtained by using the imaginary time formalism. The quasiparticle-$\pi$
mode coupling constant \picpdefn\ at finite $T$ becomes
\eqn\tempcp{
       {1\over f_\pi^2(T)}={N({\scriptstyle{\cal E}_F})\over 8
       \sqrt{{\scriptstyle{\cal E}_F}}} \int_{-\omega_D}^{\omega_D}
       {d{\scriptstyle{\cal E}}\over E^3}\sqrt{{\scriptstyle{\cal E}+
	{\cal E}_F}} \tanh {\beta \over 2}E
}
where $\beta=1/k_B T$. It is interesting to note that in the limit
${\scriptstyle{\cal E}_F}\gg\omega_D\gg\Delta$ the coupling \tempcp\ is simply
$f_\pi^2(T)=4\Delta^2(T)/N({\scriptstyle{\cal E}_F})$ as $T\rightarrow0$.
One immediate consequence of \tempcp\ is that at $T=T_c$ the integral is
divergent and so the coupling constant vanishes, i.e., $f_\pi(T_c)=0$. Hence
the determination of $T_c$ remains unaffected by the amplitude mode correction.

Similarly we can obtain the finite temperature expressions for
$\delta Z_\Delta$ and $\delta Z_\Psi$. At the Fermi surface we have
\eqn\tempmass{\eqalign{
        \delta Z_\Delta (T)=f_\pi^2(T)\,{N({\scriptstyle{\cal E}_F}) \over 8
        \sqrt{{\scriptstyle{\cal E}_F}}} \int_{-\omega_D}^{\omega_D}
	d{\scriptstyle{\cal E}}\,\sqrt{{\scriptstyle{\cal E}+{\cal E}_F}}
	\int_{-1}^1 d(\cos\theta)\,{1\over{\Delta^2-(E+E_\sigma)^2}}
        {1\over{\Delta^2-(E-E_\sigma)^2}} \cr
        \times\left[{1\over E}(\Delta^2+E^2-E_\sigma^2)\tanh{\beta E\over 2}
        +{1\over E_\sigma}(\Delta^2-E^2+E_\sigma^2)\coth{\beta E_\sigma\over 2}
        \right]}
}
\eqn\tempwaveren{\eqalign{
       \delta Z_\Psi (T)=f_\pi^2(T)\,{N({\scriptstyle{\cal E}_F}) \over 8
        \sqrt{{\scriptstyle{\cal E}_F}}} \int_{-\omega_D}^{\omega_D}
	d{\scriptstyle{\cal E}}\,\sqrt{{\scriptstyle{\cal E}+{\cal E}_F}}
	\int_{-1}^1 d(\cos\theta)\,{1\over{\Delta^2-(E+E_\sigma)^2}}
        {1\over{\Delta^2-(E-E_\sigma)^2}} \cr
       \times {1\over E_\sigma}\left[2E E_\sigma \tanh{\beta E\over 2}
       +(\Delta^2-E^2-E_\sigma^2)\coth{\beta E_\sigma\over 2}\right]}
}
where the coupling $f_\pi(T)$ is determined from \tempcp\ and
\eqn\energy{\eqalign{
	E^2 &={\scriptstyle{\cal E}}^2+\Delta^2 \cr
	E_\sigma^2&={\textstyle{8\over 3}}{\scriptstyle{\cal E}_F}^2
      	(1+\half{{\scriptstyle{\cal E}}\over{\scriptstyle{\cal E}_F}})
	-{\textstyle{8\over 3}}{\scriptstyle{\cal E}_F}^2
	\sqrt{1+{{\scriptstyle{\cal E}}\over{\scriptstyle{\cal E}_F}}}
	\cos\theta+4\Delta^2}
}
In the limit ${\scriptstyle{\cal E}_F}\gg\omega_D\gg\Delta$ and $T=0$ the
expressions \tempmass\ and \tempwaveren\ reduce to \weisstau\ and \waveren.
Combining these corrections with the finite temperature BCS gap equation,
\schrieffer\ we obtain the complete finite temperature equation for
$\Delta(T)$:
\eqn\tempgap{
	1={1\over 4 \sqrt{{\scriptstyle{\cal E}_F}}}
	VN({\scriptstyle{\cal E}_F})\int_{-\omega_D}^{\omega_D}
	{d{\scriptstyle{\cal E}} \over E}\,\sqrt{{\scriptstyle{\cal E}
	+{\cal E}_F}} \tanh{\beta \over 2}E +\delta Z_\Delta (T)
	+\delta Z_\Psi (T)
}
The solution of \tempgap\ is shown in \solnfig\ for typical values of
the parameters in a type-II superconductor. The biggest deviation is at
$T=0$ and decreases until $T=T_c$, where there is no change from the
BCS result. Such a scenario may be occurring in the organic superconductor
$({\rm BEDT}$-${\rm TTF})_2 {\rm I}_3$ where a proposal, \organicref\ to place
the gap at $6\,{\rm cm}^{-1}$ may be consistent with the observation that
the gap below $T_c$ is reduced from the BCS value ($2\Delta(T=0)=20\,
{\rm cm}^{-1}$), while $T_c$ remains unchanged.

For a thorough discussion of the realistic cases, however, one has to
take into account the complexities of the electronic band structure
and phonon spectra. One of the important effects is the mixing of the
amplitude mode with the original Coulomb and phonon interactions \lv. This
mixing is proportional to $\Delta/{\scriptstyle{\cal E}_F}$ and occurs
because of the intrinsic particle-hole asymmetry relative to the Fermi
surface. It again results in a reduction of the pairing forces and is
most significant in type-II or organic superconductors.

It should be noted that if one considers so called 'neutral'
superconductors and includes the effect of the pure Goldstone $\tau_2$
mode then one finds that the feedback is positive for the $\tau_1$
term. This will almost cancel against the negative feedback of the
amplitude mode \weisstau. However the contribution to the $\bf 1$ term
is the same sign as for the amplitude mode and will add to the
amplitude correction to give $\Delta_W^{\tau_2}=2\Delta \delta
Z_\Psi$.

The feedback effects of the collective bosonic modes on the
superconducting state may also be relevant for the recent high $\rm
T_c$ superconductors where $\Delta/{\scriptstyle{\cal E}_F} \sim 0.1$
\kresin. However without a complete understanding of the mechanism
involved in high $\rm T_c$ materials at present, we can only but
speculate on these effects.

We would like to thank M.~Weger, whose correspondence initiated this
work. This work was supported in part by NSF contract PHY-91-23780.

\listrefs
\listfigs

\vfill
\eject

\nopagenumbers

{\topinsert
        \epsfxsize=5.5in
        \epsfysize=1.2in
        \vbox{\epsfbox{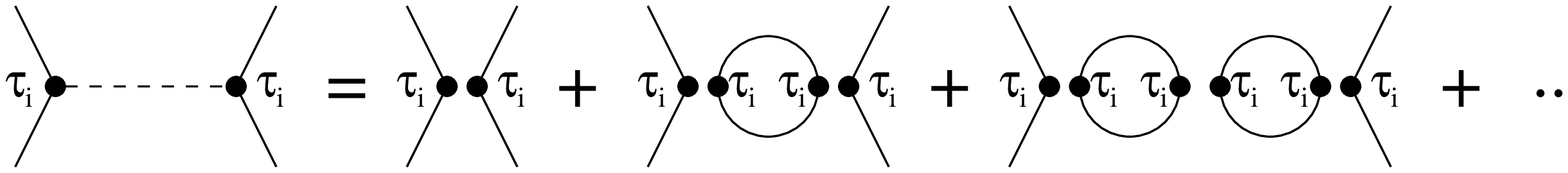}}
        \vskip 2.0in
\endinsert
}

{\topinsert
        \epsfxsize=4.5in
        \epsfysize=2.0in
        \vbox{\epsfbox{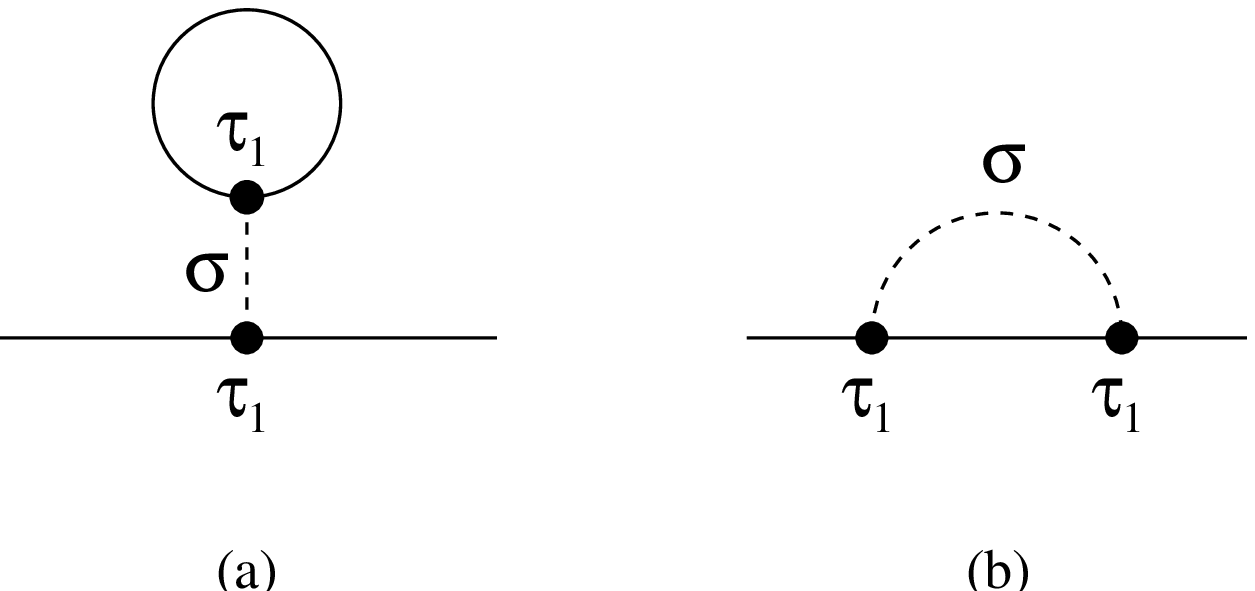}}
\endinsert
}

{\topinsert
        \epsfxsize=5.1in
        \epsfysize=7.1in
        \vbox{\epsfbox{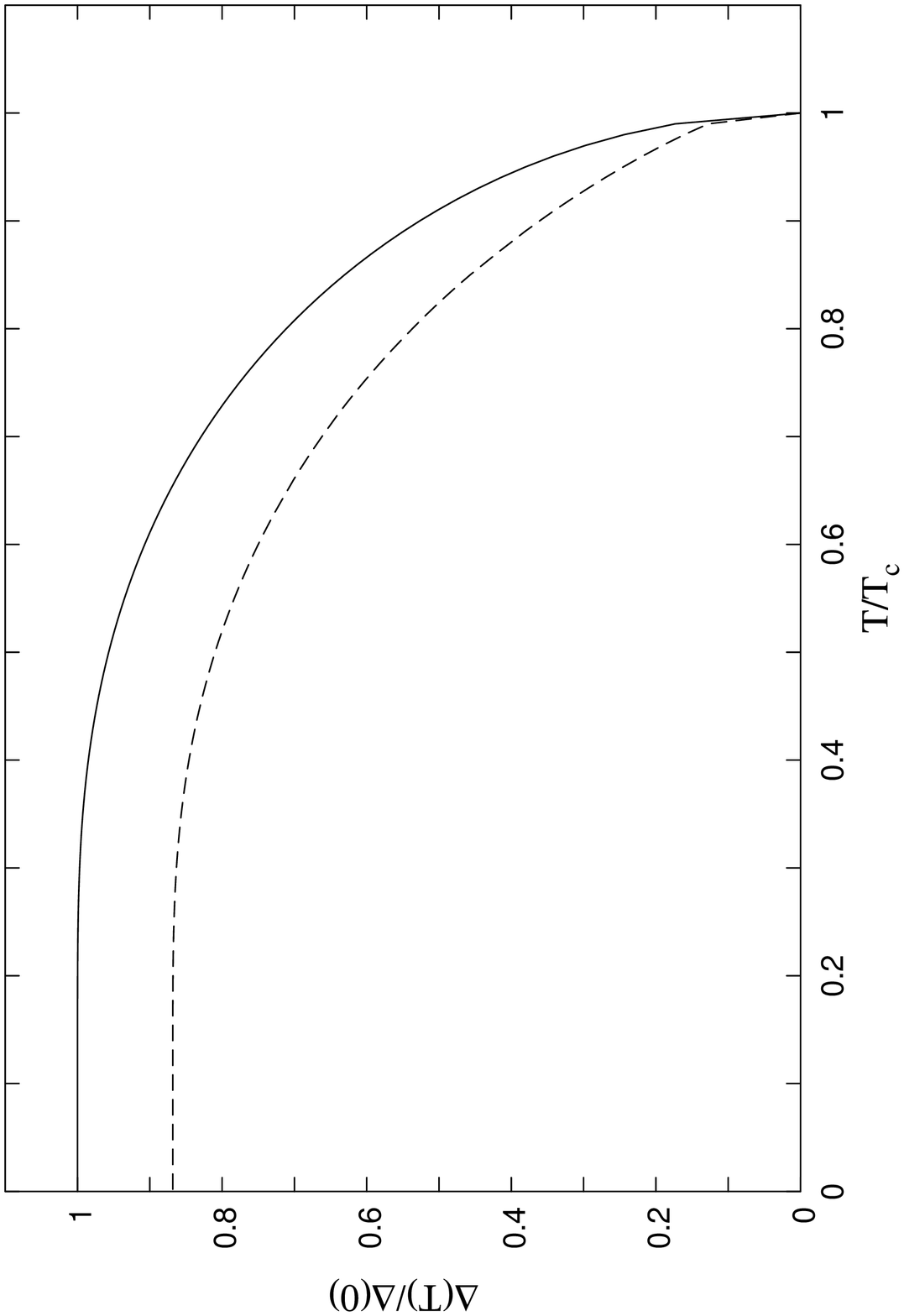}}
\endinsert
}

\bye